\documentclass[12pt]{article}
\pdfoutput=1

\usepackage[utf8]{inputenc}

\usepackage{draft}
\usepackage{putex}
 
\usepackage[all]{xy}

\usepackage{stackrel,amssymb}

\usepackage{graphicx}
\usepackage{caption}
\usepackage{amsmath,bm}
\usepackage{array}
\usepackage{subcaption}
\usepackage{epstopdf}
\usepackage{enumerate}
\usepackage{cite}
\usepackage{tensor}
\usepackage{slashed}
\usepackage[utf8]{inputenc}
\usepackage{rotating}
\usepackage{bigfoot}
\usepackage[
      colorlinks=true,
      linkcolor=blue,
      urlcolor=blue,
      filecolor=black,
      citecolor=red,
      ]{hyperref}

\usepackage{tikz-cd}

\def\XXint#1#2#3{{\setbox0=\hbox{$#1{#2#3}{\int}$ }
		\vcenter{\hbox{$#2#3$ }}\kern-.6\wd0}}

\numberwithin{equation}{section}

\newcommand{\CC}{\Gamma}

\usepackage{mathtools}

\def\<{\langle}
\def\>{\rangle}
\def\pa{\partial}
\def\ve{\varepsilon}
\def\ep{\epsilon}
\def\sD{\slashed{D}}

\def\mRP{\mathbb{RP}}

\usepackage{bm}
 \newcommand{\bD}{{\bm \Delta}}

\usepackage{fancybox}
\usepackage{cite}
\usepackage{framed}
\definecolor{shadecolor}{rgb}{0.9,0.9,0.95}
\definecolor{refkey}{rgb}{0.5,0.5,0}
\definecolor{labelkey}{rgb}{0.5,0.5,0}
\definecolor{citekey}{rgb}{0.5,0.5,0}
\definecolor{darkgreen}{rgb}{0,0.5,0}
\definecolor{darkblue}{cmyk}{0.9,0.9,0,0}
\definecolor{darkred}{rgb}{0.6,0,0.3}

\newcommand{\leftrarrows}{\mathrel{\raise.75ex\hbox{\oalign{%
				$\scriptstyle\leftarrow$\cr
				\vrule width0pt height.5ex$\hfil\scriptstyle\relbar$\cr}}}}
\newcommand{\lrightarrows}{\mathrel{\raise.75ex\hbox{\oalign{%
				$\scriptstyle\relbar$\hfil\cr
				$\scriptstyle\vrule width0pt height.5ex\smash\rightarrow$\cr}}}}
\newcommand{\Rrelbar}{\mathrel{\raise.75ex\hbox{\oalign{%
				$\scriptstyle\relbar$\cr
				\vrule width0pt height.5ex$\scriptstyle\relbar$}}}}

\makeatletter
\def\leftrightarrowsfill@{\arrowfill@\leftrarrows\Rrelbar\lrightarrows}
\newcommand{\xleftrightarrows}[2][]{\ext@arrow 3399\leftrightarrowsfill@{#1}{#2}}
\makeatother

\begin{document}

\preprint{}

	\institution{CMSA}{Center of Mathematical Sciences and Applications, Harvard University, Cambridge, MA 02138, USA}
	\institution{HU}{Jefferson Physical Laboratory, Harvard University,
		Cambridge, MA 02138, USA}

\title{
From $\mathcal{N}=4$ Super-Yang-Mills  on $\mR \mP^4$  \\
 to bosonic Yang-Mills on $\mRP^2$
}

\authors{Yifan Wang\worksat{\CMSA,\HU}}

\abstract{
We study the four-dimensional $\cN=4$ super-Yang-Mills (SYM) theory on the unorientable spacetime manifold $\mRP^4$. Using supersymmetric localization, we find that for a large class of local and extended SYM observables preserving a common supercharge $\cQ$, their expectation values are captured by an effective two-dimensional bosonic Yang-Mills (YM) theory on an $\mRP^2$ submanifold. This paves the way for understanding $\cN=4$ SYM on $\mRP^4$ using known results of YM on $\mRP^2$. As an illustration, we derive a matrix integral form of the SYM partition function on $\mRP^4$ which, when decomposed into discrete holonomy sectors, contains subtle phase factors  due to the nontrivial $\eta$-invariant of the Dirac operator on $\mRP^4$. We also comment on  potential applications of our setup for AGT correspondence, integrability and bulk-reconstruction in AdS/CFT that involve cross-cap states on the boundary.
}
\date{}

\maketitle

\tableofcontents

\pagebreak
 
 \section{Introduction}
 Quantum field theories with time reversal symmetry can be formulated on unorientable spacetime manifolds. This simple idea has lead to exciting developments in the 
 classification of topological phases and detection of subtle anomalies that involve time reversal symmetry \cite{Kapustin:2014dxa,Witten:2015aba,Hsieh:2015xaa,Seiberg:2016rsg,Witten:2016cio,Tachikawa:2016cha,Tachikawa:2016nmo,Barkeshli:2016mew}, as well as a refinement of the electric-magnetic duality in abelian gauge theories \cite{Metlitski:2015yqa}.
  However relatively little is known about the dynamics of strongly interacting theories on unorientable spacetimes beyond two dimensions.
 Luckily the bootstrap approach to conformal field theories (CFT) is well suited for this task (see \cite{Poland:2018epd} for a recent review).
 
 A familiar family of unorientable manifolds are the real projective spaces $\mRP^{d}$ of even dimensions, which are realized by a free orientation-reversing $\mZ_2$ quotient of $S^d$ (or equivalently $\mR^d\cup \{\infty\}$ by a Weyl transformation) and 
 preserve a large residual (Euclidean) conformal subalgebra $\mf{so}(d+1)$ similar to the case of co-dimension one defects (planar or spherical) in flat space.\footnote{The boundary CFT can be thought of as defined by a  $\mZ_2$ quotient of the flat space with a co-dimension one fixed loci.} Consequently one can formulate a bootstrap program for the basic observables in the CFT on $\mRP^{d}$, namely the correlation function of local operators, similar to the case with a domain wall or boundary defect \cite{Liendo:2012hy,Billo:2016cpy}. Putting the CFT on $\mRP^d$ introduces new observables beyond those on the flat space, given by the one-point functions of \textit{normalized} scalar primary operators $\cO(x)$ \cite{Nakayama:2016cim},
 \ie
 \la \cO(x)  \ra={ h_\cO \over (1+  x^2)^{\Delta_\cO}}\,
 \fe
where the position dependence is fixed by the conformal dimension $\Delta_\cO$ due to the residual symmetry (which also requires the one-point function of spinning primaries to vanish). The coefficients $h_\cO$ furnish the basic structure constants for the CFT on $\mRP^d$.  Along with the OPE of local operators, they determine general correlation functions on $\mRP^d$. Solving the CFT on $\mRP^d$ amounts to fixing these coefficients $h_\cO$ in terms of the ordinary OPE data of the CFT on $\mR^d$ by exploring constraints from the (residual) conformal symmetry, crossing symmetry and unitarity,\footnote{Note that unlike the more familiar four-point function bootstrap, here the combinations of OPE coefficients that appear in the conformal block decomposition (e.g. of two-point functions on $\mRP^d$) have no obvious positivity properties.} possibly supplemented by additional dynamical inputs from other methods. This program has been pursued for the Ising model in two and three dimensions via numerical techniques \cite{Nakayama:2016cim}, and for Lee-Yang theory in $6-\ep$ dimensions \cite{Hasegawa:2016piv} and Wilson-Fisher theory in $4-\ep$ dimensions \cite{Hasegawa:2018yqg} to leading orders in the $\ep$-expansion.\footnote{See \cite{Hogervorst:2017kbj} for a reformulation of the bootstrap equations on $\mRP^d$.} 

Gauge theories in four-dimensions offer a rich playground to advance this program. On one hand, a large class of CFTs are produced by  renormalization group (RG) flows from four-dimensional Yang-Mills theories coupled to matter. On the other hand, on a topologically nontrivial manifold such as $\mRP^4$, the gauge theory observables become sensitive to fine details of the theory, such as global structures of the gauge group, topological couplings, and the spectrum of extended defects \cite{Witten:1998wy,Aharony:2013hda,Razamat:2013opa,Gaiotto:2014kfa,Gukov:2014gja,Metlitski:2015yqa}. Due to strong coupling effects, few observables in general four-dimensional gauge theories can be obtained analytically beyond perturbation theory. Fortunately in supersymmetric gauge theories, the supersymmetric localization method \cite{Nekrasov:2002qd,Pestun:2007rz,Pestun:2016zxk} allows for extractions of exact results for a large subset of the correlation functions. When combined with the bootstrap program, it provides a powerful way to solve the CFT. This has been particularly successful in the study of the four-dimensional $\cN=4$ super-Yang-Mills (SYM) theory on flat space, with possibly extended conformal defects (see for example \cite{Liendo:2016ymz,Giombi:2017cqn,Giombi:2018qox,Liendo:2018ukf,Giombi:2018hsx,Binder:2019jwn,Chester:2019pvm,Chester:2019jas,Chester:2020dja,Drukker:2020swu}).

 In this note, we initiate the study of the four-dimensional $\cN=4$ super-Yang-Mills theory on $\mR\mP^4$ using an extension of the localization setup of \cite{Pestun:2009nn,Wang:2020seq}. There, a particular supercharge $\cQ$ in the  superconformal algebra  $\mf{psl}(4|4)$\footnote{The Lorentzian $\cN=4$ superconformal algebra is  $\mf{psu}(2,2|4)$ which is a real form of the complex Lie superalgebra $\mf{psl}(4|4)$.  Here we study the CFT in the Euclidean signature obtained from a Wick rotation. As usual, one then loses the reality condition on the fermionic generators of the superalgebra. This is not a problem for our analysis if the theory is invariant under the supersymmetry transformations regardless any reality conditions, which is the case here \cite{Pestun:2007rz}. For the same reason, we will also not impose reality conditions on the fermionic generators in the Euclidean superconformal subalgebra $\mf{osp}(4|4)$ on $\mRP^4$ in this paper.} was used to a localize the theory on $S^4$ (resp. $HS^4$) to two-dimensional (constrained) Yang-Mills theory on a great $S^2$ (resp. $HS^2$). Since the antipodal quotient of $S^4$ gives the $\mRP^4$ (with round metric), one naturally expects that by implementing a  supersymmetric  $\mZ_2$ identification, the SYM on $\mRP^4=S^4/\mZ_2$ should lead to the bosonic YM on $\mRP^2=S^2/\mZ_2$ upon localization. Indeed as will see, such an identification exists and the $\cN=4$ SYM can be defined on $\mRP^4$ preserving a half-BPS subalgebra $\mf{osp}(4|4)$ which contains the supercharge $\cQ$. The partition function of the YM theory on $\mRP^2$ has a simple combinatorial formula in terms of the representation data of the gauge group \cite{Witten:1991we}, which in turn determines the partition function of the $\cN=4$ SYM on $\mRP^4$, which can be re-expressed into a single matrix model. Thanks to the general discussions in \cite{Wang:2020seq}, observables of the $\cN=4$ SYM on $\mRP^4$, involving local operators as well as defects preserving the common supercharge $\cQ$, translate to (defect) observables in the YM theory on $\mRP^2$, which can be further reduced to computations in the relevant matrix model as illustrated in \cite{Wang:2020seq,Komatsu:2020sup}. We leave the detailed investigation of such observables to a future publication.     

The rest of the paper is organized as follows. In Section~\ref{sec:localization}, we identify the supersymmetric $\mZ_2$ involution that defines $\cN=4$ SYM on $\mRP^4$ preserving the $\mf{osp}(4|4)$ subalgebra and carry out the localization computation with respect to the aforementioned supercharge $\cQ$  that leads to the two dimensional YM on $\mRP^2$. In Section~\ref{sec:2dym}, we derive a matrix integral form for the SYM partition function on $\mRP^4$ using two dimensional gauge theory techniques and compare with results from an alternative localization procedure discussed in \cite{Bawane:2017gjf,LeFloch:2017lbt}. We also comment on subtle phase factors that have been missing thus far in  a gluing formula for the $\mRP^4$ partition function.
  We end by a brief summary and discuss a number of future directions in Section~\ref{sec:conclusion}.

\section{Localization of the $\cN=4$ SYM on $\mRP^4$}
\label{sec:localization}

 \subsection{The SYM on $\mRP^4$ from involution}
 \label{sec:inv}
 On flat space $\mR^4$ with coordinates $x_\m$, the real projective space $\mR\mP^4$ is defined by identifying points related by a fixed-point-free involution 
 \ie
 \iota: x_\m \to x'_\m\equiv -{x_\m \over x^2}\,.
 \label{iotaflat}
 \fe
 The Jacobian of the transformation is
 \ie
 {\pa  x'^\m \over \pa x^\n}  =-x'^2 I_{\m\n} (z),\quad 
 \fe
 where
 \ie
 I_{\m\n}=\D_{\m\n}-{2x_\m x_\n \over  x^2} ,~\quad I_{\m}{}^{\n} I_{\n \rho}=\D_{\m \rho }\,.
 \fe
The Jacobian has negative determinant since $\det I_{\m\n}=-1$  
 \ie
\det \left({\pa  x'^\m \over \pa x^\n} \right)=-|x'|^8=-{1\over |x|^8}
 \fe
and consequently $\mRP^4=\mR^4/\mZ_2^\iota$ is unorientable.
 
The residual conformal symmetry on $\mRP^4$ is generated by rotations $M_{\m\n}$ and the combination of translation and special conformal transformations $P_\m -K_\m$. Together they defines the subalgebra
 \ie
 \mf{so}(5)\subset \mf{so}(5,1)\,.
 \fe
 
 The $\cN=4$ SYM theory enjoys the superconformal symmetry $\mf{psl}(4|4)$ on flat space which includes the conformal symmetry and R-symmetry
 \ie
 \mf{psl}(4|4) \supset \mf{so}(5,1)\times \mf{so}(6)_R\,,
 \fe 
 as bosonic subalgebras, as well as Poincaré and conformal supercharges that are conveniently packaged into a conformal Killing spinor
 \ie
 \ve =\ep_s + x^\m \tilde\CC_\m \ep_c 
 \label{flatsc}
 \fe 
 where $\ep_s$ and $\ep_c$ are constant 16 component spinors of $Spin(10)$.\footnote{We adopt the convention of \cite{Pestun:2009nn} for the spinors and gamma matrices. The 10d chiral and anti-chiral gamma matrices are $\CC_M$ and $\tilde \CC_M$ respectively which satisfy the Clifford algebra $\{\CC_M,\tilde\CC_N\}=2\D_{MN} 1_{16}$. $\ep_s$ and $\ep_c$ are chiral and antichiral 16-component spinors. See next section for more details on the notation.} We would like an extension of the involution $\iota$ on  $\mf{psl}(4|4)$ such that a half-BPS subalgebra is preserved. In general, the action of $\iota$ on the supercharge takes the form\footnote{For the action of $\iota$ on general spinor fields see \eqref{priid}. Note that $\ve$ has conformal weight $-{1\over 2}$.}
 \ie
 \iota_{\rm SYM}:~ \ve(x) \to  i x^\m \tilde\CC_\m \cR \ve(x')
 \fe
 where $\cR$ is induced by an (outer)automorphism of the $\mf{so}(6)_R$ symmetry.  The preserved supercharges are those satisfying
 \ie
 ix^\m\tilde\CC_\m \cR \ve(x')= \ve(x)\,.
 \label{iotasol}
 \fe
 Consistency of the superconformal algebra requires $\ve \CC_\m \ve$ to transform under $\iota$ as a vector. Combined with the requirement of solutions to \eqref{iotasol}, we conclude
 \ie
 \cR=-  \CC_{790}\,,
 \label{Ronsp}
 \fe
 up to an $\mf{so}(6)_R$ rotation. The preserved 16 supercharges  are parametrized by constant spinors satisfying
\begin{shaded}
\ie
\ep_c= -i \CC_{790} \ep_s 
\label{RP4sc}
\fe\end{shaded}
\noindent
which generate a half-BPS subalgebra of $\mf{psl}(4|4)$,
 \ie
 \mf{osp}(4|4) \supset \mf{so}(4,1)\oplus \mf{so}(3)_{568}\oplus\mf{so}(3)_{790}\,.
 \fe 

We note that this subalgebra includes the supercharge $\cQ$ of \cite{Pestun:2009nn} (we follow the convention of \cite{Wang:2020seq} here) that defines the 2d Yang-Mills sector  on $S^2$ in the SYM which is specified by additional projectors (only three are independent) as below,
\begin{shaded}
 \ie
 	\Gamma_{1890}\ep_s=\Gamma_{1279}\ep_s=\Gamma_{1370}\ep_s=\Gamma_{2390}\ep_s=-\Gamma_{1238}\ep_s=\ep_s,\quad \ep_c=-i   \Gamma_{790}   \ep_s
\,.
 \label{Qspinor}
 \fe  
\end{shaded} 
In the next section, we will exploit this fact to show that upon identification by $\iota_{\rm SYM}$, the resulting 2d sector of the SYM on $\mRP^4$ is given by the YM on $\mRP^2=S^2/\mZ_2^\iota$.

In general, CFTs on real projective spaces are defined  via the identification induced by $\iota$ on the primary operators on the flat space. For scalar, vector and fermion operators of dimension $\Delta$, such identifications take the following form\footnote{The spinors here are defined on $\mRP^4$ with a ${\rm pin}^+$ structure ($\iota^2=1$ acting on the fermions in the Euclidean signature). There are two pin$^+$ structures on $\mRP^4$ related by a sign flip in the identification of the fermions at points $x$ and $x'$ related by the involution $\iota$ on the covering space  (see Appendix~A of \cite{Witten:2015aba} for a detailed review and also  \cite{Kapustin:2014dxa,Metlitski:2015yqa,Witten:2016cio,Guo:2017xex}  for relevant discussions).}
\ie
\phi(x')=|x|^{ 2\Delta}\phi (x),\quad V_\m(x')= -|x|^{ 2\Delta} I_{\m} {}^{\n}V_\n (x),\quad \psi (x')=  i|x|^{2\Delta-1}{ \C_\m x^\m  }\psi (x)\,.
\label{priid}
\fe
Here to define the $\cN=4$ SYM on $\mRP^4$ preserving the half-BPS supersymmetry, we require the following supersymmetric identification due to $\iota_{\rm SYM}$
\ie
\Phi_I(x')=|x|^2 (\cR\Phi)_I (x ),\quad A_\m(x')= -|x|^{2} I_{\m} {}^{\n}A_\n (x ),\quad \Psi (x')=-i |x|^{2}{ \tilde \CC_\m x^\m}\cR\Psi (x )\,,
\label{idR4}
\fe
where $\cR$ is (induced) by \eqref{Ronsp}. 
 More explicitly,
\ie
 (\cR\Phi)_I=(-\Phi_5,-\Phi_6,\Phi_7,-\Phi_8,\Phi_9,\Phi_0)\,.
 \label{Ronphi}
\fe
Note that the identification for the fermions are completely fixed by that of the bosons (either $\Phi_I$ or $A_\mu$) by compatibility with the supersymmetry $\ve$ \eqref{RP4sc} that we want to preserve (see \eqref{SUSYos} in the next section).

 By stereographic map from the flat space, we put the (Euclidean) CFT on the sphere $S^4$ with metric
 \ie
 ds^2=e^{2\Omega} dx^2,\quad e^\Omega\equiv {2R\over 1+ x^2}\,.
 \label{S4metric}
 \fe
  Then the inversion $\iota$ in \eqref{iotaflat} simply amounts to the antipodal map on the $S^4$. In particular, the two hemispheres $HS^4_\pm$ for $|x|>1$ and $|x|<1$ are identified point-wise, and the equator $S^3$ at $|x|=1$ is reduced to $\mRP^3$ (a special case of the lens space). The residual (superconformal) symmetry on $\mRP^4$  with round metric \eqref{S4metric} is simply given by the Weyl transformation of the flat space counterparts.  In particular the conformal Killing spinors become
   \ie
   \varepsilon_{S^4}= e^{{1\over 2}\Omega} \varepsilon 
   \label{ksS4}
   \fe
   with constant spinors $\ep_{s,c}$ subject to the same constraint \eqref{RP4sc}.  The antipodal map clearly commutes with the $\mf{so}(5)_{\rm rot}$ symmetry of the $S^4$ and this is identified with the residual conformal symmetry on $\mRP^4$ (with round metric).
    The SYM fields are also related by  
\ie
 A_\m \to A_\m  \,,\quad \Phi_I\to \Phi_I e^{-\Omega}\,,\quad \Psi\to \Psi e^{-{3\over 2}\Omega} 
 \label{R4toS4}
\fe
thanks to the Weyl symmetry.

 Before we end this section, let us make some comments in relation to the the AdS/CFT correspondence \cite{Maldacena:1997re,Witten:1998qj}. The holographic dual of the $\cN=4$ SYM in the large $N$ limit is given by the IIB string theory on AdS$_5\times S^5$ background with metric
  \ie
  ds^2={dx_\m dx^\m  +dy^I dy^I \over |y|^2} \,,
  \label{AdS5S5m}
  \fe
  and self-dual five-form flux
   \ie
   F_5={4\over L} \left({\rm vol}_{AdS_5}+{\rm vol}_{S^5} \right) \,.
   \label{flux}
   \fe
   The $\cN=4$ superconformal symmetry is realized in the bulk by Killing spinors on AdS$_5\times S^5$ 
   \ie
   \ve_{\rm AdS}(x,y)={1\over |y|} (\ep_s+ (x^\m \Gamma_\m+ y^I \Gamma_I)\ep_c)\,,
   \fe
   and they are related to the conformal Killing spinor \eqref{flatsc} on the boundary by taking the asymptotic limit
   \ie
   \lim_{|y| \to 0}|y|\ve_{\rm AdS} = \ve\,.
   \fe
The boundary involution $\iota_{\rm SYM}$ corresponds to an orbifold by  an $\mZ_2$ isometry in the bulk
\ie
\iota_{\rm SYM}:~(x_\m, y^I)\to \left(-{x_\m \over {|x|^2+|y|^2}},  {1\over {|x|^2+|y|^2}}(\cR y)^I \right)\,,
\label{idiib}
\fe
where the action on the internal coordinates $y^I$ follows from that on the SYM scalars $\Phi_I$ in \eqref{Ronphi}. Note that this $\mZ_2$ preserves the orientation of the 10d spacetime while the individual volume forms on the AdS$_5$ and $S^5$ factors are odd. Thus to preserve the five-form flux background \eqref{flux}, the involution $\iota_{\rm SYM}$ in the IIB string theory needs to be supplemented by the worldsheet parity $\Omega_{\rm ws}$. In other words, the IIB background dual to the supersymmetric $\mRP^4$ is given by an orientifold of the usual AdS$_5\times S^5$ background. Moreover, $\iota_{\rm SYM}$ has a fixed locus  located at the center of the AdS$_5$ given by $x_\m=0$, and wrapping a internal $S^2\subset S^5$ given by
\ie
y_{5}=y_6=y_8=0,~y_7^2+y_9^2+y_0^2=1\,.
\fe

\subsection{The supersymmetric action for $\cN=4$ SYM on $\mRP^4$}
In this section, we present the action of the $\cN=4$ SYM on $\mRP^4$ preserving off-shell supersymmetry, obtained by performing a $\mZ_2$ identification of the SYM fields by $\iota_{\rm SYM}$ on the covering space $S^4$.

We start by reviewing the action for 4d $\cN=4$ SYM with gauge group $G$ on $S^4$ \cite{Berkovits:1993hx,Pestun:2007rz},
\ie
S_{\rm SYM} = -{1\over 2 g_4^2}\int_{S^4} d^4 x\, \sqrt{g} \tr \Bigg(
{1\over 2}F_{MN}F^{MN}-\Psi {\Gamma}^M D_M\Psi  +{2\over R^2} \Phi^I \Phi_I -K^m K_m
\Bigg)\,.
\label{SYMos}
\fe
Here we follow the convention in \cite{Pestun:2009nn} to regroup the 4d spacetime index $\m=1,2,\dots,4$ and $\mf{so}(6)_R$ symmetry index $I=5,6,\dots,9,0$ into the 10d index $M=1,2,\dots,9,0$. Correspondingly the 10d gauge fields $A_M$ contains the 4d gauge fields $A_\m$ and adjoint scalars $\Phi_I$. The gaugino $\Psi$ transforms as a chiral spinor of $Spin(10)$.  
Note that $\Gamma_\m=e_{\hat\m \m}\bm\Gamma^{\hat \m}$, where $e_{\hat\m}^\m$ is the vielbein and  $\bm\Gamma^{\hat M}$ denotes flat space 10d Gamma matrices in the chiral basis (we will not distinguish between $\bm\Gamma^{\hat I}$ and $\Gamma^{ I}$ for Gamma matrices in the internal directions).
 $K_m$ with $m=1,\dots 7$ are auxiliary fields which serve to give an off-shell realization of the supercharge that we will use to localize the theory. 
The gauge covariant derivative is defined by $D\equiv d+A$ with curvature $F=dA+A\wedge A$. In terms of its $\mf{g}$ components, $A_M\equiv A_M^a T_a$ comes with real coefficients $A_M^a$ and \textit{anti-hermitian} generators $T_a$ of  $\mf{g}$. The trace $\tr(\cdot,\cdot)$ is the Killing form of $\mf{g}$ and is related to the usual trace in a particular representation $\lambda$ by $\tr={1\over 2T_\lambda}\tr_\lambda$, where $T_\lambda$ denotes the Dynkin index of $\lambda$. For $\mf{g}=\mf{su}(N)$, this is identical to the trace in the fundamental representation $\tr=\tr_F$. Finally the generators $T^a$ are normalized by $\tr(T_a T_b)=-{1\over 2}\D_{ab}$. We set the four-dimensional theta angle $\theta=0$ in this paper.\footnote{Another possible value of the \textit{bare} theta angle compatible with the orientation-reversing $\mZ_2$ quotient is $\theta=\pi$.}

The superconformal transformation of the SYM fields are 
\ie
&\D_\ve A_M= \varepsilon \Gamma_M \Psi  \,,
\\
&\D_\ve\Psi ={1\over 2}F_{MN}\Gamma^{MN}\varepsilon+{1\over 2}\Gamma_{\m I}\Phi^I \nabla^\m \varepsilon+K^m \n_m \,,
\\
&\D_\ve K^m=-\n^m\Gamma^M D_M\Psi\,,
\label{SUSYos}
\fe
where the conformal Killing spinor $\varepsilon$ is given by $\varepsilon_{S^4}$ in \eqref{ksS4} but we have dropped the subscript to avoid clutter in notations.
 In order for the subalgebra generated by $\D_\ve$ to close off-shell, the auxiliary 10d chiral spinors $\n^m$ with $m=1,\dots,7$ in \eqref{SUSYos} are chosen to satisfy
\ie
\varepsilon \Gamma^M \n_m=0\,,~~ \n_m\Gamma^M\n_n=\D_{mn}\varepsilon \Gamma^M \varepsilon\,,~~\n^m_\A \n^m_\B +\ep_\A \ve_\B ={1\over 2}\ve \CC_M \ve \tilde \CC^M_{\A\B}\,.
\label{PSpinor}
\fe
Consequently $\D^2_\ve$ gives rise to a combination of the bosonic symmetries (including an $SO(7)$ rotation of $K_m$) of the SYM action \eqref{SYMos}.
 
The SYM on $\mRP^4$ with the round metric is defined by implementing the following identifications between the SYM fields at antipodal points $x$ and $x'$ on the $S^4$,
\begin{shaded}
	\ie
\Phi_I(x')=  (\cR\Phi)_I (x),\quad A_\m(x')= - I_{\m} {}^{\n}A_\n (x),\quad \Psi (x')=-i  {  {\tilde {\bm\CC}_\m} x^\m\over |x|}\cR\Psi (x),\quad K_m(x')=  K_m (x)
\,.
\label{idS4}
\fe
\end{shaded}
\noindent
which follows from the identification \eqref{idR4} on $\mR^4$ after taking into account the Weyl factors in \eqref{R4toS4}. We emphasize that this identification is a symmetry of the action \eqref{SYMos}, and respects the off-shell SUSY transformation laws \eqref{SUSYos}, provided that we implement the same identification for the auxiliary pure spinors as in \eqref{iotasol},
\ie
\nu_m(x')=-i x^\m \tilde{\bm\CC}_\m \cR \nu_m(x)\,,
\fe
which is compatible with the pure spinor constraints in \eqref{PSpinor}.\footnote{In checking these one may find useful the following identity for gamma matrices $x^\n x^\rho {\bm \Gamma}_\n  {\bm\Gamma}_\m {\bm\Gamma}_\rho =-  |x|^2 I_{\m\n} {\bm\Gamma}^\m $.}

\subsection{Localization to 2d YM on $\mRP^2$}
We now specialize to the supercharge $\cQ$ defined by the Killing spinor $\ve$ satisfying \eqref{Qspinor}. It generates an $\mf{su}(1|1)$ subalgebra\footnote{We have suppressed an $SO(7)$ rotation which only acts on the auxiliary scalars $K_m$.}
\ie
\cQ^2=- 2(
M_\perp - R_{56}
)
\fe
where $M_\perp \equiv {1\over 2}(K_4-P_4)$ generates rotation transverse to a distinguished great $S^2$ in the $S^4$, which we denote by $S^2_{\rm YM}$
\ie
S^2_{\rm YM}:~x_4=0,~x_1^2+x_2^2+x_3^2=1\,,
\fe
and $R_{56}$ is an R-symmetry rotation (e.g. it acts on the scalars $\Phi_{5,6}$).

As explained in \cite{Pestun:2009nn} (see also \cite{Wang:2020seq} for a streamlined version), 
upon turning on a $\cQ$-exact (and $\cQ$-closed) deformation (the localizing term)  as
\ie
S_{\rm SYM} \to S_{\rm SYM} + t \int_{S^4} d^4x \sqrt{g}  \tr \D_\ve (\bar\Psi\D_\ve \Psi)\,,
\fe
the SYM path integral on $S^4$ localizes to the BPS locus $\D_\ve  \Psi=0$ in the limit $t\to \infty$. For $\cQ$-closed observables in the SYM, including the partition function of the SYM on $S^4$, such a deformation doesn't affect their expectation values, therefore we are free to take this limit. A careful analysis of the BPS locus shows that the smooth BPS configurations are uniquely determined by the value of an emergent $G$ gauge field $\cA$
 \ie
 \cA=&A+i \epsilon_{ijk} x^j \phi^k dx^i\,,
 \fe
 on the $S^2_{\rm YM}$
 with
 \ie 
  (\phi_1,\phi_2,\phi_3)\equiv (\Phi_7,\Phi_9,\Phi_0)\,.
  \fe
Furthermore, they are weighted by an effective 2d Yang-Mills action
 \ie
 S_{\rm YM} \equiv -{1\over  g_{\rm YM}^2}\int_{S^2}   dV_{S^2} \tr (\star \cF)^2\,,
 \fe
 with the 2d YM coupling $g_{\rm YM}$ related to the 4d SYM coupling $g_4$ by
 \ie
 g_{\rm YM}^2=-{g_4^2\over 2\pi R^2}\,.
 \label{2d4d}
 \fe
An important distinction between this effective 2d theory and the usual 2d Yang-Mills is that the (unstable) instanton contributions (nontrivial solutions to $D\star \cF=0$ parametrized by the cocharacter lattice of $G$)  \cite{Witten:1992xu,Blau:1991mp,Blau:1993hj,Minahan:1993tp} are excluded in \cite{Pestun:2009nn}. For this reason, we refer to this theory as the constrained-Yang-Mills theory (cYM).\footnote{In the original paper \cite{Pestun:2009nn}, this was named ``almost Yang-Mills'' or aYM.} 

To summarize, we have the following from \cite{Pestun:2009nn}
 \begin{equation*}
 \begin{tikzcd}[column sep=6pc]
 \text{4d $\cN=4$ $G$ SYM on $S^4$}   \arrow{r}{\cQ-{\rm localization}} &
 \text{2d $G$ cYM on $S^2_{\rm YM}$}
 \end{tikzcd}
 \end{equation*}

Now the localization of the SYM on $\mRP^4=S^4/\mZ_2$ follows verbatim from the derivations in \cite{Pestun:2009nn,Wang:2020seq} since all BPS equations are compatible with the identification \eqref{idS4} by construction. The end result simply amounts to the identification \eqref{idS4} at the level of the 2d cYM theory on the $S^2_{\rm YM}$, 
\ie
P_{ij}\cA_j(-\vec x)=-P_{ij}\cA_j(\vec x)\,,
\fe
where $P_{ij}\equiv \D_{ij}-x_i x_j$ is the projector to the tangent space of $S^2_{\rm YM}$. The 2d YM action  is simply
\ie
S_{\rm YM} \equiv -{ 1\over  g_{\rm YM}^2}\int_{\mRP^2}   dV_{S^2} \tr (\star \cF)^2\,.
\fe
Therefore we conclude
\begin{shaded}
 \begin{equation*}
 \begin{tikzcd}[column sep=6pc]
 \text{4d $\cN=4$ $G$ SYM on $\mRP^4$}   \arrow{r}{\cQ-{\rm localization}} &
 \text{2d $G$  YM on $\mRP^2_{\rm YM}$}
 \end{tikzcd}
 \end{equation*}
with
\ie
\mRP^2_{\rm YM}:~\{x_4=0,~x_1^2+x_2^2+x_3^2=1\}/(x_i\sim -x_i)\,.
\label{RP2YM}
\fe
\end{shaded} 
\noindent
Note that since the nontrivial (unstable) instantons are forbidden on $\mRP^2$, there is no difference between constrained and ordinary 2d YM in this case.

\section{2d YM on $\mRP^2$  and a new matrix model}
\label{sec:2dym}

The 2d Yang-Mills theory on a Riemann surface $\Sigma$ enjoys a large spacetime symmetry given by ${\rm SDiff}(\Sigma)$, namely  the area-preserving diffeomorphisms on $\Sigma$. This makes the theory rather rigid (almost like a 2d TQFT) and consequently its partition function  on general $\Sigma$ has a simple combinatorial description  in terms of representation theory data of the underlying gauge group $G$ \cite{Migdal:1975zg,Witten:1991we,Witten:1992xu,Blau:1991mp,Blau:1993hj,Cordes:1994fc}. 

For example, the partition function of the standard 2d YM on $S^2$ is given by a weighted sum over all irreducible representations $\lambda$ of $G$ 
\ie
Z^{\rm YM}_{S^2}=\sum_\lambda d_\lambda^2 e^{-  \pi R^2g_{\rm YM}^2 c_2(\lambda)}\,,
\label{S2YM}
\fe
where $d_\lambda$ and $c_2(\lambda)$ denote the dimension and second Casimir of the representation $\lambda$. 
For $\mRP^2$, we have instead\footnote{See \cite{Stanford:2019vob} for a different representation of the YM partition function on $\mRP^2$ computed for gauge group $PGL(2,\mR)$ using the Reidemeister-Ray-Singer torsion. }
\ie
Z^{\rm YM}_{\mRP^2}=\sum_\lambda d_\lambda  f_\lambda e^{-  {1\over 2}\pi R^2g_{\rm YM}^2 c_2(\lambda)}\,,
\fe
where $f_\lambda$ is the Frobenius-Schur indicator for the representation $\lambda$
\ie
f_\lambda =\begin{cases}
	1 &  \lambda ~{\rm is~real}\,,
	\\
	-1 &  \lambda ~\text{is~pseudo-real}\,,
	\\
	0 &  \lambda ~{\rm is~complex}\,.
\end{cases}
\label{FS}
\fe
Via our localization argument in the last section, they give the partition functions of the $\cN=4$ SYM on $S^4$ and $\mRP^4$ respectively where the 2d and 4d gauge couplings are related as in \eqref{2d4d}. Indeed, the $S^4$ partition function of SYM is given by a Gaussian matrix model via a different localization computation in \cite{Pestun:2007rz} which agrees with the 2d YM answer \eqref{S2YM} after projection to the zero-instanton sector and taking into account appropriate counter-terms \cite{Wang:2020seq}. 

The partition function of the $\cN=4$ SYM on $\mRP^4$ is however unknown.  Here  in Section~\ref{sec:RP2pf} we use the effective 2d YM description to determine a matrix integral expression for the $\mRP^4$ partition function with interesting features.  Then in Section~\ref{sec:RP2cp} we comment on the relation between our matrix integral to  important partial results for the localization of $\cN=2$ gauge theories on $\mRP^4$ \cite{LeFloch:2017lbt,Bawane:2017gjf} using the $\cN=2$ supercharge of \cite{Pestun:2007rz}.

\subsection{Partition function }
\label{sec:RP2pf}

For simplicity we focus on the case with $G=U(N)$. We label the representations  of $U(N)$ by an $N$-tuple of strictly decreasing integers $(\ell_1,\dots,\ell_N)$  as in \cite{Wang:2020seq}. The dimension and quadratic Casimir of the representation $\lambda$ are  given by
\ie
d_\lambda={\Delta(\ell_i) \over \Delta(N-i)},\quad c_2(\lambda)=-{N(N^2-1)\over 12}+\sum_{i=1}^N\left(\ell_i-{N-1\over 2}\right)^2\,.
\fe
In particular  the fundamental representation has $(\ell_1,\dots,\ell_N)=(N,N-2,N-3,\dots,0)$ and $c_2(F)=N$. Here $\Delta$ is the Vandermonde determinant defined for $N$-tuples $x_i$ as,
\ie
\Delta(x_i)\equiv \prod_{1\leq i< j\leq N} (x_i-x_j)\,.
\fe 
To obtain the $S^2$ partition function of the cYM, one needs to implement a projection to the zero-instanton sector of the full YM partition function \eqref{S2YM}. This gives, after subtracting certain counter-terms \cite{Wang:2020seq}, 
\ie
Z^{\rm cYM}_{S^2} =&{ 1\over N!  }\int \prod_{i=1}^N da_i\,  \Delta^2(a_i)e^{-    {8\pi^2\over g_4^2} \sum_{i=1}^Na_i^2 } \,,
\label{ZS2}
\fe
which agrees with the expected partition function of the $\cN=4$ SYM on $S^4$ \cite{Pestun:2007rz}. 

For the cYM on $\mRP^2$, as explained in the previous section, there is no need to do the zero-instanton projection. Thus the partition function for the $U(N)$ Yang-Mills is simply
\ie
Z^{\rm YM}_{\mRP^2}=&
 {e^{  {N(N^2-1) \over 24} \pi R^2g_{\rm YM}^2} } \sum^{\rm real}_{\ell_i \in \mZ} \left( {\Delta(\ell_i) \over \Delta(N-i)} \right)e^{-{1\over 2}  \pi R^2g_{\rm YM}^2  \sum_{i=1}^N\left(\ell_i-{N-1\over 2}\right)^2}\,,
\label{RP2sum}
\fe
where the sum is over real representations of $U(N)$ labeled by $\ell_1>\ell_2>\dots>\ell_N$. If we write $\ell_i=\lambda_i-i+N$, then $\sum_{i=1}^N \lambda_i$ determines the $U(1)$ charge of the representation and $[\lambda_1-\lambda_2,\dots,\lambda_{N-1}-\lambda_N]$ gives the Dynkin label for the corresponding $SU(N)$ representation. The representation $\lambda$ is complex unless the $U(1)$ charge vanishes $\sum_{i=1}^N \lambda_i=0$ and the $SU(N)$ Dynkin label is left-right mirror-symmetric (i.e. $\lambda_1-\lambda_2= \lambda_{N-1}-\lambda_N$ etc). Together, they imply that the real representations are given by 
\ie
\lambda_i+\lambda_{N+1-i}=0\,,
\fe 
or equivalently
\ie
\ell_i +\ell_{N+1-i}=N-1\,.
\fe
 
To rewrite the $\mRP^2$ partition function of the 2d YM  into a matrix model we use the following Fourier transformation formula \cite{Minahan:1993tp,Wang:2020seq}
\ie
   &\int  \left(\prod_{i=1}^N dz_i \right)\Delta(z_i) e^{2\pi i \sum_{i=1}^N z_i (\ell_i-{N-1\over 2})}  e^{-  {2 \pi  \over R^2g_{\rm YM}^2}  \sum_{i=1}^Nz_i^2}
   \\
   =&
 i^{N(N-1)\over 2} \left(   R^2 g_{\rm YM}^2\over 2\right)^{N^2\over 2}  \Delta(\ell_i)  e^{-{1\over 2}  \pi R^2g_{\rm YM}^2  \sum_{i=1}^N\left(\ell_i-{N-1\over 2}\right)^2}\,,
\fe
and  \eqref{RP2sum}  becomes
\ie
Z^{\rm YM}_{\mRP^2}=&
{e^{  {N(N^2-1) \over 12} \pi R^2g_{\rm YM}^2} \over   G(N+1) } i^{N(1-N)\over 2} \left(   R^2 g_{\rm YM}^2\over 2\right)^{-{N^2\over 2}} 
\\
&\times  \sum^{\rm real}_{\ell_i \in \mZ} \int  \left(\prod_{i=1}^N dz_i \right)\Delta(z_i) e^{2\pi i \sum_{i=1}^N z_i (\ell_i-{N-1\over 2})}  e^{-  {2 \pi  \over R^2g_{\rm YM}^2}  \sum_{i=1}^Nz_i^2}\,.
\fe
Taking into account the 2d counter-terms as in \cite{Wang:2020seq} (noting that the Euler characteristic $\chi(\mRP^2)=1$), we have the renormalized action,  which by abusing the notation we will still denote by $Z^{\rm YM}_{\mRP^2}$,
 \ie
 Z^{\rm YM}_{\mRP^2}=&
  i^{{N(1-2N) \over 2}}   \sum^{\rm real}_{\ell_i \in \mZ} \int  \left(\prod_{i=1}^N dz_i \right)\Delta(z_i) e^{2\pi i \sum_{i=1}^N z_i (\ell_i-{N-1\over 2})}  e^{-  {2 \pi  \over R^2g_{\rm YM}^2}  \sum_{i=1}^Nz_i^2}\,.
 \fe
Parametrizing the $U(N)$ holonomy (up to conjugation) as $U=( e^{i z_1},\dots, e^{i z_N})$, the character for the representation $\lambda$ is
\ie
\chi_{\lambda}(e^{i z_i})={\det_{ij} e^{i z_i l_j}\over \det_{ij} e^{i z_i (N- j)}},~~i,j=1,2,\dots,N\,.
\fe
We can thus write after some algebra
\ie
Z^{\rm YM}_{\mRP^2}=&
  {i^{N(1-2N)\over 2} }  \sum_{\lambda  }^{\rm real}\int  \left(\prod_{i=1}^N dz_i \right)\Delta(z_i) {\Delta(e^{2\pi i z_i})}\chi_\lambda(e^{2\pi i z_i}) 
  e^{-\pi i (N-1)\sum_i z_i} e^{-  {2 \pi  \over R^2g_{\rm YM}^2}  \sum_{i=1}^Nz_i^2}\,.
  \label{RP2pft}
\fe

We will use the following identity
\begin{shaded}
\ie
\sum_{\lambda  }^{\rm real} \chi_\lambda(e^{2\pi i z_i})
=& \sum_{m_i\in \{0,1\}}  {\Delta(e^{ \pi i (z_i+m_i)}  )\Delta(e^{ -\pi i (z_i+m_i)}  )\over \Delta (e^{ 2\pi i z_i}) \Delta (e^{ -2\pi i z_i})}\,.
\label{impid}
\fe
\end{shaded}
This can be derived as follows
 \ie
\sum^{\rm real}_{\lambda  }\chi_\lambda(e^{2\pi i z_i})=&{1\over 2^N N!}\left(\prod_{i=1}^N \int_0^2 d y_i \right)\Delta(e^{ \pi i y_i})  \Delta(e^{ -\pi i y_i})  \sum_{\lambda}\chi_\lambda(e^{2\pi i y_i})\chi_\lambda(e^{2\pi i z_i})
\\
=& \sum_{m_i\in \{0,1\}}  {\Delta(e^{ \pi i (z_i+m_i)}  )\Delta(e^{ -\pi i (z_i+m_i)}  )\over 2^N \Delta (e^{ 2\pi i z_i}) \Delta (e^{ -2\pi i z_i})}
 \fe
where  in the first equality we have used the following expression for the Frobenius-Schur indicator
\ie
f_\lambda=\int dU \chi_\lambda(U^2)\,,
\fe
in terms of an integral of the character with respect to the Haar measure on $G$, and further reduced the integral to the maximal torus of $G$ using the Weyl integral formula \cite{Blau:1993hj}. In the second step we have used the completeness relation \cite{Witten:1991we,Blau:1991mp}, 
  \ie
  \sum_{\lambda}\chi_\lambda(e^{2\pi i z_i})\chi_\lambda(e^{2\pi i y_i})={N!\D_p(z_i+y_i) \over \Delta  (e^{2\pi i z_i})\Delta  (e^{-2\pi i z_i})} \,.
  \fe

Applying  \eqref{impid} to \eqref{RP2pft}, we obtain
\ie
Z^{\rm YM}_{\mRP^2}=&
{i^{N(1-2N)\over 2}\over 2^N N!}  \int  \left(\prod_{i=1}^N dz_i \right) {\Delta(z_i)\over \Delta(e^{-2\pi i z_i})}
e^{-\pi i (N-1)\sum_i z_i} e^{-  {2 \pi  \over R^2g_{\rm YM}^2}  \sum_{i=1}^Nz_i^2}
 \sum_{m_i\in \{0,1\}}  \Delta(e^{ \pi i (z_i+m_i)}  )   \Delta(e^{ -\pi i (z_i+m_i)}  ) 
\fe
Note that $g_{\rm YM}^2<0$ for the effective 2d YM from the localization of the 4d SYM \eqref{2d4d}. Thus the appropriate integral contour requires a Wick rotation
\ie
a_j =  i z_j\,,
\fe
consequently we have
\begin{shaded}
 \ie
Z^{\rm SYM}_{\mRP^4}= Z^{\rm YM}_{\mRP^2}=&
 {i^{-{N(N+2)\over 2}}\over 2^N N!}  \int  \left(\prod_{i=1}^N da_i \right) {\Delta(a_i)\over \bD(  a_i)}e^{-  {4 \pi^2  \over g_4^2}  \sum_{i=1}^N a_i^2} \sum_{m_j\in \{0,1\}}  \bD\left(a_j+i m_j   \over 2\right)^2\,,
 \label{RP2mm}
   \fe
   \end{shaded}
   \noindent
where we have defined
\ie
\bD(a_i)\equiv \prod_{i<j} 2\sinh (\pi (a_i-a_j))\,.
\fe
The generalization to arbitrary gauge group $G$ is clear which we state without proof (we have also dropped an overall phase factor below),\footnote{See next section for additional evidence for this expression from a gluing construction of the SYM partition function on $\mRP^4$.} 
\begin{shaded}
	\ie
	Z^{\rm SYM}_{\mRP^4}= Z^{\rm YM}_{\mRP^2}=
	{1 \over 2^r |W_G|}  \int_{\mf{t}}  d^r {\bm a}\, {\Delta_G({\bm a})\over \bD_G( {\bm a})}e^{-  {4 \pi^2  \over g_4^2} \tr {\bm a}^2} \sum_{ {\bm m}\in \Lambda_{\rm cochar}^G/2\Lambda_{\rm cochar}^G }  \bD_G\left({\bm a}+i {\bm m}  \over 2\right)^2\,.
	\fe
\end{shaded}
\noindent
where $r$ denotes the rank of $G$, $W_G$ is the Weyl group and $d^r {\bm a}$ is the standard $W_G$-invariant measure on the Cartan subalgebra $\mf{t}\subset \mf{g}$. The generalized Vandermonde factors are products over the positive roots of $G$
\ie
\Delta_G({\bm a})\equiv \prod_{\A \in \Delta(G)^+}  \A ({\bm a}),\quad 
\bD_G({\bm a})\equiv \prod_{\A \in \Delta(G)^+} 2\sinh \,(\pi  \A ({\bm a}))\,.
\fe
  The sum over ${\bm m}$ is determined by the cocharacter lattice $\Lambda_{\rm cochar}^G$ which depends on the global structure of $G$ (e.g. the fundamental group and center of $G$ are given by
  	 $\pi_1(G)=\Lambda_{\rm cochar}^G/\Lambda_{\rm coroot}^\mf{g}$ and $Z(G)=\Lambda_{\rm coweight}^\mf{g}/\Lambda_{\rm cochar}^G$).  Recall the $S^4$ partition function of the SYM \cite{Pestun:2007rz} does not depend on such global structures, which sets  apart the SYM on $\mRP^4$.

\subsection{Discrete holonomy sectors and phase factors}
\label{sec:RP2cp}

In \cite{Bawane:2017gjf,LeFloch:2017lbt}, the localization of $\cN=2$ gauge theories on $\mRP^4$ were studied in relation to the AGT correspondence \cite{Alday:2009aq,Alday:2009fs,Alday:2010vg,Drukker:2010jp}  where the dual Toda theory observables involve certain boundary or cross-cap states. The localizing supercharge in these studies follows from the one in \cite{Pestun:2007rz} which we will call $\cQ_{\cN=2}$. Importantly, this supercharge, in contrary to the $\cQ$ we have been using, squares to a combination of isometry and R-symmetry generators that admit a single fixed point on $\mRP^4$ (a pair of antipodal fixed points on the covering space $S^4$).

More explicitly, as explained in \cite{Wang:2020seq}, the localizing supercharge $\cQ$ that defines the 2d YM sector of the 4d $\cN=4$ SYM decomposes as
\ie
\cQ =\cQ_{\cN=2}^+ +\cQ_{\cN=2}^-\,,
\fe
into a pair of supercharges $\cQ_{N=2}^\pm$  in isomorphic $\cN=2$ subalgebras $\mf{osp}(4|2)_\pm \subset \mf{osp}(4|4)\subset \mf{psl}(4|4)$ defined by the projections
\ie
\CC_{5670} \ep_s= \pm \ep_s,\quad \ep_c= -i\CC_{790}\ep_s\,.
\fe
These $\cN=2$ supercharges satisfy
\ie 
&\{\cQ^+_{\cN=2},\cQ^+_{\cN=2}\}=-M_\perp-M_{13}+R_{56}-R_{70}\,,
\\
&\{\cQ^-_{\cN=2},\cQ^-_{\cN=2}\}=-M_\perp+M_{13}+R_{56}+R_{70}\,,
\\
&\{ \cQ^+_{\cN=2},\cQ^-_{\cN=2}\}=0\,,
\fe
and they are are precisely of the type considered in \cite{Pestun:2007rz} to localize $\cN=2$ gauge theory on $S^4$ to a zero-dimensional matrix model.

Consequently, by localizing a general $\cN=2$ theory on $\mRP^4$ with the supercharge $\cQ_{\cN=2}$ ($\cQ_{\cN=2}^\pm$ for the $\cN=4$ SYM), one expects to arrive directly at a matrix model similar to that of \eqref{RP2mm} where the precise form of the integrand will depend on the matter content of the 4d gauge theory through their one-loop determinants around the BPS locus. A novelty of the $\mRP^4$ computation compared to the $S^4$ case is an additional discrete label for the BPS solutions, by the holonomy of the gauge fields along the nontrivial one-cycle $H_1(\mRP^4,\mZ)=\mZ_2$. In \cite{Bawane:2017gjf}, the one-loop determinants for $\cN=2$ vector- and hyper-multiplets in the trivial holonomy sector on $\mRP^4$ were computed explicitly.
 Meanwhile \cite{LeFloch:2017lbt} provided a complementary perspective on the $\mRP^4$ partition function by gluing the partition function of the 4d $\cN=2$ theory on a hemisphere $HS^4$ \cite{Bullimore:2014nla,Gava:2016oep} with half-BPS Dirichlet boundary conditions (that are compatible with the identification \eqref{idS4} restricted to the equator $S^3$),  and the lens space $S^3/\mZ_2$ partition function \cite{Gang:2019juz,Benini:2011nc,Alday:2012au,Imamura:2012rq,Imamura:2013qxa} of the 3d boundary modes which fall into 3d $\cN=2$ multiplets thanks to the residual supersymmetry.\footnote{See  \cite{Dedushenko:2018tgx} for general discussions on gluing constructions of supersymmetric partition functions on oriented spacetime manifolds.} For general $\cN=2$   gauge theories on $\mRP^4$ with the round metric, the partition function takes the following form
 \ie
 Z_{\mRP^4}(g_4^2)= {1\over |W_G|}\int_{\mf t} d^r {\bm a}   \sum_{ {\bm m}\in \Lambda_{\rm cochar}^G/2\Lambda_{\rm cochar}^G } \underbrace{ Z_{S^3/\mZ_2}(\bm{a},\bm{m})}_{\substack{\text{one-loop } \\ \text{determinants on $S^3/\mZ_2$}}} \underbrace{\hat Z_{HS^4}^{\rm Dir}(\bm{a},\bm{m},g_4^2)}_{\substack{\text{$HS^4$ wavefunction} \\ \text{with Dirichlet b.c.}}}   \,,
 \label{gluing}
 \fe
 where $\bm m$ labels the discrete holonomy of the $G$ gauge fields (up to the Weyl group). The BPS boundary conditions of the $\cN=2$ gauge theory on $HS^4$ are specified by a constant value of the vector multiplet scalar $\phi=\bm{a}\in \mf{t}$\footnote{In the $\cN=4$ SYM, $\varphi$ corresponds to the combination $\Phi_9+i\Phi_8$ in our convention here.} and a flat connection $A \in \mf{t}$ on $S^3$.\footnote{With the Dirichlet boundary condition, the gauge transformations approach identity at the boundary $S^3$.} We expect the dependence of the hemisphere partition function $\hat Z_{HS^4}^{\rm Dir}(\bm{a},A,g_4^2)$ on the flat connections to be  a phase. This is because gluing two copies of $HS^4$ with opposite orientations (the partition functions related by complex conjugation) gives the partition function on $S^4$ which admits no nontrivial flat connections. Focusing on the case when the flat connections are labelled by $\mZ_2$ holonomies between antipodal points on the boundary $S^3$,\footnote{Such connections are picked out by the localization of the boundary modes on $S^3/\mZ_2$.} we have
 \ie
 \hat Z_{HS^4}^{\rm Dir}(\bm{a},\bm{m},g_4^2)= e^{i\varphi(\bm{m})} Z_{HS^4}^{\rm Dir}(\bm{a},g_4^2)   \,,
 \label{dirphase}
 \fe
where $Z_{HS^4}^{\rm Dir}(\bm{a},g_4^2)$ is the usual hemisphere partition function with $ \bm{m}=0$ \cite{Bullimore:2014nla,Gava:2016oep}
and  $e^{i\varphi(\bm{m})}$ is an undetermined phase factor.\footnote{In \cite{LeFloch:2017lbt}, this phase factor was treated separately and fixed for the $\cN=2$ $SU(N)$ conformal SQCD by explicitly comparing to the cross-cap wavefunction in the $A_{N-1}$ Toda theory.}

Focusing now on the $\cN=4$ SYM with $U(N)$ gauge group, the boundary modes of the 4d $\cN=4$ multiplet on $HS^4$ relevant for the gluing construction organize into one 3d $\cN=4$ vector multiplet \cite{Dedushenko:2018tgx}. In terms of the 3d $\cN=2$ subalgebra, they correspond to one vector multiplet and one chiral multiplet. The lens space one-loop determinants of general 3d $\cN=2$ multiplets were determined in \cite{Gang:2019juz,Benini:2011nc,Alday:2012au,Imamura:2012rq,Imamura:2013qxa}. Here the $\cN=2$ chiral multiplet is in the adjoint representation of $G$ and has  $U(1)_r$ charge $r=1$, thus contributes trivially to the one loop determinant \cite{Benini:2011nc}. The nontrivial contribution is entirely due to the $\cN=2$ vector multiplet, which gives 
\ie
Z_{S^3/\mZ_2}(a_i,m_i)=\bD\left(a_j+i m_j   \over 2\right)\bD\left(a_j-i m_j   \over 2\right)\,,
\fe
for a fixed $\mZ_2$-valued $U(N)$ holonomy labelled by $m_i$. This is closely related to the last factor in \eqref{RP2mm} by 
\ie
\sum_{m_j\in \{0,1\}}  \bD\left(a_j+i m_j   \over 2\right)^2
=
\sum_{m_j\in \{0,1\}} e^{{\pi i \over 2} \sum_{i\neq j}(m_i-m_j)^2} \bD\left(a_j+i m_j   \over 2\right)\bD\left(a_j-i m_j   \over 2\right)\,.
\fe
Together with  the hemisphere partition function of the $\cN=4$ SYM with Dirichlet boundary condition   \cite{Bullimore:2014nla,Gava:2016oep,Dedushenko:2018tgx,Wang:2020seq} (up to an overall $a_i$ independent factor) and trivial $m_i$
\ie
Z_{HS^4}^{\rm Dir}(a_i,g_4^2)= {i^{-{N(N+2)\over 2}}\over 2^N N!}  {\Delta(a_i)\over \bD(  a_i)}e^{-  {4 \pi^2  \over g_4^2}  \sum_{i=1}^N a_i^2}\,,
\fe
we see the two expressions for the SYM matrix model on $\mRP^4$ from 2d YM on $\mRP^2$ \eqref{RP2mm} and the gluing (factorization) formula \eqref{gluing} agree if the phase factor in \eqref{dirphase} is given by
\ie
e^{i\varphi(m_i)}=\prod_{i\neq j} e^{{\pi i \over 2} (m_i-m_j)^2}\,.
\fe 
For general $G$, a similar manipulation gives a product over its roots, 
 \ie
e^{i\varphi( \bm{m})}=\prod_{\A \in \Delta} e^{{\pi i \over 2}  \A({\bm m})^2}\,.
\label{genphase}
 \fe  
For simple $G$ (e.g. $SU(N)$), this phase factor is trivial since $\sum_{\A \in \Delta}\A({\bm m})^2 = 2 h^\vee (G)\tr ({\bm m}^2) $ where $h^\vee(G)$ is the dual Coxeter number. The triviality of the phase factor extends in an obvious way to semi-simple $G$ and only fails when $G$ contains $U(1)$ factors.

It would be interesting to derive this phase factor from first principle by a careful analysis following the localization setup of \cite{LeFloch:2017lbt,Bawane:2017gjf}. Here we offer some observations for why this is natural. 
We consider the weak coupling limit $g_4^2\to 0$ of the hemisphere partition function $\hat Z_{HS^4}^{\rm Dir}(\bm{a},\bm{m},g_4^2)$. In this case, we expect the nontrivial phase factor to come from the fermion one-loop determinant since the bosonic contributions are manifestly positive.  
The $\cN=4$ vector-multiplet contains two 4d Dirac fermions in the adjoint representation of the gauge group $G$. 
The supersymmetric boundary condition that defines  $\hat Z_{HS^4}^{\rm Dir}(\bm{a},\bm{m},g_4^2)$ sets half of the fermions in the $\cN=4$ SYM to zero on the boundary $S^3$ \cite{Gava:2016oep,Dedushenko:2018tgx,Wang:2020seq}. Together with the antipodal identification \eqref{idS4} on the boundary $S^3$, it follows that the relevant fermion determinant (that depends on $\bm m$) is effectively the determinant of the Dirac operator $\det (\sD_{\bm m}) $ on $\mRP^4$ for a single fluctuating Dirac fermion $\psi$ in the adjoint representation of $G$. As explained in \cite{Witten:2015aba,Metlitski:2015yqa,Witten:2016cio}, the determinant $\det (\sD_{\bm m})$ contributes a nontrivial phase depending on the $\mZ_2$-holonomy $e^{\pi i \A(\bm {m})}$ along the one-cycle of $\mRP^4$ (dropping $\bm m$ independent factors below),
  \ie
\det (\sD_{\bm{m}}) \propto |\det (\sD_{\bm{m}})| \prod_{\A \in \Delta(G) }e^{-2\pi i \eta (\A({\bm m})) }
 \fe
in terms of the $\eta$-invariant of a Dirac fermion on $\mRP^4$ which takes the following values (in the convention of \cite{Metlitski:2015yqa})
\ie
\eta \equiv \pm {1\over 8}
\fe
respectively for the two pin$^+$ structures on $\mRP^4$. Now the pin$^+$ structures associated to the Dirac fermion $\psi$ along the root vector $\A$  
 are interchanged depending on whether the $\mZ_2$-holonomy $e^{\pi i \A(\bm {m})}=\pm 1$ is even or odd. Thus its $\eta$-invariant can be written as,
 \ie
e^{-2\pi i \eta\A({\bm m}) } =e^{-{2\pi i \over 8}}e^{{2\pi i \over 4} \A({\bm m})^2 }\,.
 \fe
 Taking products over the root vectors $\A$ of $G$, we obtain precisely the phase factor \eqref{genphase} found earlier.
  
\section{Conclusion and discussion}
\label{sec:conclusion}

In this note, we have initiated the study of the 4d $\cN=4$ super-Yang-Mills on unorientable spacetime $\mRP^4$ using supersymmetric localization. By extending the previous works \cite{Pestun:2009nn,Wang:2020seq}, we found that the $\cN=4$ SYM on $\mRP^4$ localizes to 2d YM on  $\mRP^2$. Invoking known results about the partition function of 2d YM on general Riemann surfaces, we derived a matrix model expression for the partition function of the $\cN=4$ SYM on $\mRP^4$.  We also commented on relations to other localization results on $\mRP^4$ using a different supercharge \cite{Bawane:2017gjf,LeFloch:2017lbt} and pointed out subtle phase factors in a gluing formula for the partition function that come from  $\eta$-invariant of the fermions on $\mRP^4$.

There are a number of interesting directions to be explored further and we  discuss some of them below.
 \paragraph{Correlation functions} 
The localizing supercharge $\cQ$ here defined by \eqref{Qspinor} is precisely the one considered in \cite{Wang:2020seq}, with respect to which general ${1\over 16}$-BPS defect observables were classified. This means our localization setup of the $\cN=4$ SYM on $\mRP^4$ here can be decorated by such $\cQ$-preserving defect networks, which lead to particular insertions in the 2d YM on $\mRP^2$ following the dictionary in \cite{Wang:2020seq} and can be computed using matrix model techniques as illustrated in \cite{Wang:2020seq,Komatsu:2020sup}. In particular, this will allow us to determine correlation functions of ${1\over 8}$-BPS local operators
\ie
\cO_J(x_i)\equiv \tr (x_1 \Phi_7 +x_2 \Phi_9+x_3\Phi_0+i \Phi_8)^J
\fe
restricted to the submanifold $\mRP^2_{\rm YM}\subset \mRP^4$ \eqref{RP2YM}, possibly in the presence of a ${1\over 2}$-BPS interface defect along the equator $S^3/\mZ_2\subset \mRP^4$ at $x_1=0$.

\paragraph{AGT} As discussed in \cite{LeFloch:2017lbt,Bawane:2017gjf}, the vanilla AGT correspondence \cite{Alday:2009aq,Alday:2009fs,Alday:2010vg,Drukker:2010jp} between observables of the Toda theory on a closed and oriented Riemann surface $\Sigma$ and certain 4d $\cN=2$ theory on $S^4$ can be naturally extended to include boundaries and cross-caps on the Toda side
 by considering the 6d $(2,0)$ theory on supersymmetric quotient geometries $(S^4\times \Sigma)/\mZ_2$. If we take $\Sigma$ to be a (punctured) square torus with holomorphic coordinate $z=z_1+iz_2$ subjected to the identification $z\sim z+1 \sim z+\tau$ with $\tau={4\pi i \over g_4^2}$ and a particular degenerate primary $V_\A$ of the Toda theory inserted at $z=0$ (see \cite{Wang:2020seq} for a short review), the 6d $(2,0)$ theory of type $G$ reduced on such a $\Sigma$ gives the $\cN=4$ SYM with gauge group $G$. Furthermore if we 
  choose the 6d $\mZ_2$ quotient to act as the antipodal map on $S^4$,  according the general discussion in \cite{LeFloch:2017lbt,Bawane:2017gjf}, the $\cN=4$ SYM on $\mRP^4$ will be related to the corresponding Toda theory on $\Sigma/\mZ_2$, based on the two fundamental domains of $(S^4\times \Sigma)/\mZ_2$. The action of $\mZ_2$  on $\Sigma$ (that preserves the puncture at $z=0$) is then restricted to be $z_1 \to -z_1$ (or $z_2 \to -z_2$ related by an S-transformation). Consequently the quotient $\Sigma/\mZ_2$
 gives an annulus with a modulus ${\rm Im\,}\tau$ and a Toda degenerate primary operator inserted on one of the two boundary circles. Note that as discussed in \cite{LeFloch:2017lbt}, one expects the relevant Toda boundary states on the annulus to be  \textit{unusual} in the sense that their wavefunctions are given by a cross-cap state in the CFT. It would be interesting to have a more complete understanding of the AGT correspondence in this case which will also teach us about the $SL(2,\mZ)$ properties of the $\cN=4$ SYM on unorientable spacetime.

\paragraph{Integrability }  As previously mentioned, the $\cN=4$ SYM on $\mRP^4$ preserves the same symmetry $\mf{osp}(4|4)$ (up to a Wick rotation) as in the case with a half-BPS boundary or interface defect. Consequently we expect the integrability methods \cite{Jiang:2019xdz,Jiang:2019zig,Komatsu:2020sup} that have been successful in bootstrapping the interface one-point functions of non-BPS operators in the planar $\cN=4$ SYM to also apply here. It would be interesting to carry out this analysis for the one-point functions of $\cN=4$ SYM on $\mRP^4$.

	\paragraph{AdS/CFT}
	In the context of AdS/CFT correspondence \cite{Maldacena:1997re,Witten:1998qj}, since the bulk extension  \eqref{idiib}  of the boundary involution defining the $\mRP^4$ manifold fixes the center of AdS$_5$, one expects the holographic dual of the CFT on $\mRP^4$ involves a local \textit{bulk} operator (with gravitational dressing) inserted at the center. This falls into the general bulk-reconstruction program from the CFT (see \cite {DeJonckheere:2017qkk,Harlow:2018fse} for reviews). But in contrary to the Hamilton-Kabat-Lifschytz-Lowe (HKLL) approach \cite{Hamilton:2006az} which relies on a semi-classical gravity dual and ${1\over N}$ expansions, the identification between CFT cross-cap states ($\mRP^d$) and bulk operators are intrinsically non-perturbative and background independent. Such an identification has been explored extensively for AdS$_3/$CFT$_2$ \cite{Verlinde:2015qfa,Miyaji:2015fia,Nakayama:2015mva,Nakayama:2016xvw,Goto:2016wme,Lewkowycz:2016ukf}. In particular, correlation functions involving bulk operators in \textit{pure} gravity on AdS$_3$ are reproduced by CFT correlators with cross-cap Ishibashi states \cite{Miyaji:2015fia,Nakayama:2015mva,Nakayama:2016xvw,Goto:2016wme,Lewkowycz:2016ukf}. For higher dimensional CFTs, it is generally less clear whether such an identification between cross-cap CFT states and bulk operators is complete due to the dynamical gravity in the bulk. Nonetheless here we have seen an effective reduction of the 4d $\cN=4$ SYM on $\mRP^4$ to 2d YM on $\mRP^2$ for certain supersymmetric observables. It would be interesting to understand the implications for the bulk-reconstruction program involving boundary cross-cap states in the $\cN=4$ SYM.

\section*{Acknowledgements}
The author thanks Douglas Stanford, Sergio Benvenuti and especially Bruno Le Floch for many useful correspondences. The author is also grateful to Bruno Le Floch and Xinan Zhou for helpful comments on the draft.
This work  is  supported in part by the Center for Mathematical Sciences and Applications and the Center for the Fundamental Laws of Nature at Harvard University.

\bibliographystyle{JHEP}
\bibliography{defREF,SYMdefect,locREF}

\end{document}